\documentclass[final,3p,times,authoryear]{elsarticle}

\usepackage{amssymb}
\usepackage{amsmath}

\usepackage{url,hyperref,lineno}
\usepackage{multirow} 
\usepackage{upgreek}
\usepackage{makecell}

\begin{document}

\begin{frontmatter}



\title{3D Cardiac Anatomy Generation Using Mesh Latent Diffusion Models} 



\author[1]{Jolanta Mozyrska} 
\author[2]{Marcel Beetz} 
\author[3]{Luke Melas-Kyriazi}
\author[2]{Vicente Grau}
\author[2,4]{Abhirup Banerjee\fnref{senior}}
\author[1]{Alfonso Bueno-Orovio\corref{cor1}\fnref{senior}}
\ead{alfonso.bueno@cs.ox.ac.uk}

\fntext[senior]{These authors share senior authorship.}
\cortext[cor1]{Corresponding author.}

\affiliation[1]{organization={Department of Computer Science, University of Oxford},
            city={Oxford},
            postcode={OX1 3QD}, 
            country={United Kingdom}}

\affiliation[2]{organization={Institute of Biomedical Engineering, Department of Engineering Science, University of Oxford},
            city={Oxford},
            postcode={OX3 7DQ}, 
            country={United Kingdom}}

\affiliation[3]{organization={Visual Geometry Group, Department of Engineering Science, University of Oxford},
            city={Oxford},
            postcode={OX1 3PH},
            country={United Kingdom }}

\affiliation[4]{organization={Division of Cardiovascular Medicine, Radcliffe Department of Medicine, University of Oxford},
            city={Oxford},
            postcode={OX3 9DU},
            country={United Kingdom}}

\begin{abstract}
\noindent Diffusion models have recently gained immense interest for their generative capabilities, specifically the high quality and diversity of the synthesized data. However, examples of their applications in 3D medical imaging are still scarce, especially in cardiology. Generating diverse realistic cardiac anatomies is crucial for applications such as \textit{in silico} trials, electromechanical computer simulations, or data augmentations for machine learning models. In this work, we investigate the application of Latent Diffusion Models (LDMs) for generating 3D meshes of human cardiac anatomies. To this end, we propose a novel LDM architecture -- MeshLDM. We apply the proposed model on a dataset of 3D meshes of left ventricular cardiac anatomies from patients with acute myocardial infarction and evaluate its performance in terms of both qualitative and quantitative clinical and 3D mesh reconstruction metrics. The proposed MeshLDM successfully captures characteristics of the cardiac shapes at end-diastolic (relaxation) and end-systolic (contraction) cardiac phases, generating meshes with a 2.4\% difference in population mean compared to the gold standard.
\end{abstract}






\end{frontmatter}



\section{Introduction}

Diffusion models have recently attracted significant attention in the deep learning community and beyond due to their excellent generative capabilities, specifically the high quality and diversity of the produced data. Prominent examples of diffusion models generating 2D images from text include Stable Diffusion, built upon the work by \citet{rombach2021ldm}, and Google’s Imagen \citep{saharia2022photorealistic}. Beyond 2D spaces, recent innovations like the MeshDiffusion model \citep{liu2023meshdiffusion} have showcased their potential for generating 3D shapes like cars, chairs, and other common-life objects.

Lately, diffusion models have been applied to medical imaging as well. According to the first survey on diffusion models in medical imaging \citep{survey2023}, there has been an exponential growth in the number of publications with applications spanning from segmentation and anomaly detection to 2D and 3D image generation. Examples include Latent Diffusion Models (LDMs) \citep{rombach2021ldm}, which outperform generative adversarial network-based methods \citep{goodfellow2014gan} for generating class-conditional 2D chest X-ray images \citep{Packhauser2023}. LDMs have also demonstrated the ability to generate realistic synthetic 3D brain images \citep{pinaya2022brain}.

Building on this momentum, we aim to utilize LDMs for generating  3D meshes of cardiac anatomies. Such models can address the problem of limited annotated medical data for machine learning models. Existing studies such as \citep{chen2021synthetic} on histology images have demonstrated that supplementing real samples with synthetic data enhances the classification performance. The generated synthetic data can also be applied for electrophysiological computer simulations for \textit{in silico} trials \citep{li2022deep,li2024towards,CAMPS2024103108,CAMPS2025103361}.

While mitigating the challenges of data scarcity, it is essential to ensure the diversity of generated medical data. Given that human cardiac anatomies exhibit significant variations between individuals, diverse representations are vital. Furthermore, different aspects of this diversity align with different diseases, enabling clinicians to make precise diagnoses and treatments. Thus, employing models that can capture the broad spectrum of data diversity is paramount. The variational autoencoder based Mesh VAE \citep{beetz2022interpretable} has previously been successful in modeling diverse cardiac anatomy. In this study, we investigate the following question: \emph{can latent diffusion models be used for capturing anatomical variability and generating realistic 3D meshes of human left ventricles (LVs)?}

To this end, we introduce a novel architecture -- MeshLDM -- a latent diffusion model designed to generate 3D synthetic cardiac anatomy meshes. In this regard, our contributions include: (i) development of MeshLDM, a LDM architecture specifically designed for generating 3D meshes of cardiac anatomies; (ii) evaluation of MeshLDM's abilities to generate 3D meshes of the human LV at both the end-diastolic (relaxation) and end-systolic (contraction) cardiac phases, using a dataset of patients with acute myocardial infarction; (iii) assessing MeshLDM's ability to capture fine-detailed anatomical changes caused by cardiac disease, including changes in basal plane tilt, mid-cavity diameter, and elongation; and (iv) demonstration of MeshLDM's capability to successfully learn characteristics of end-diastolic and end-systolic meshes, including differences in myocardial thickness. Altogether, and to the best of the authors' knowledge, this represents the first study that leverages LDMs for generating 3D meshes of cardiac anatomies.

The rest of the article is organized as follows:
in Section~\ref{s:methods}, we describe the architecture of the proposed MeshLDM and its training process. Next in Section~\ref{s:results}, we present both qualitative and quantitative analyses of the generated meshes. Finally in Section~\ref{s:discussion}, we discuss the results of our experiments, address their limitations, and propose future improvements and extensions.

\section{Materials and methods}
\label{s:methods}
\subsection{Overview}
In this study, we introduce a novel architecture -- MeshLDM -- a latent diffusion model designed to generate 3D meshes of cardiac anatomies (Figures~\ref{fig:ldm_ae} and \ref{fig:ldm_dm}). We train and evaluate our model using a dataset initially designed for the study of post-myocardial infarction (MI) based on magnetic resonance imaging (MRI) \citep{beetz2022interpretable}. It comprises of meshes representing human LVs, as detailed in Section~\ref{s:dataset}. We describe the architecture of the proposed MeshLDM in Section~\ref{s:mesh_ldm_architecture} and explain the training process of the MeshLDM, in particular the autoencoder and the denoising model, in Section~\ref{s:network_training}.

\subsection{Dataset}
\label{s:dataset}
We use the dataset as described in \citep{beetz2022interpretable} along with the same pre-processing steps.

The dataset consists of 1034 meshes of human LVs at the end-diastolic (ED) cardiac phase and 1034 meshes at the end-systolic (ES)  phase \citep{CORRALACERO20221563, CorralAcero2024}. The data was acquired from post-MI patients and includes both ST-elevation MI (STEMI) and non-ST-elevation MI (NSTEMI) patients. The acquired cine MR images were processed to obtain 3D surface meshes of the LVs \citep{Lamata2014JRSI}.
We split the mesh dataset into approximately 70\% training, 5\% validation, and 25\% test datasets.
Before inputting the data into the MeshLDM model, we apply standardization (subtracting the mean and dividing by the standard deviation) to each mesh. Each mesh has the same number of consistently ordered vertices, allowing us to treat the vertex coordinates as a fixed-size feature vector. We standardize each coordinate $(x, y, z)$ by subtracting the mean and dividing by the standard deviation computed across the training dataset.

We train two separate models – one for ED data and another for ES data. The meshes from ED and ES phases vary considerably in shape. If we naively combined the ED and ES datasets and trained a single model, it would attempt to learn an average representation of both cardiac phases, likely resulting in the generation of anatomically unrealistic shapes blending the ED and ES characteristics (e.g., a partially contracted ventricle with a large volume). Such shapes would lack clinical utility because they would not accurately represent either phase of the cardiac cycle.

\subsection{MeshLDM architecture}
\label{s:mesh_ldm_architecture}
\begin{figure*}[!t]\centering\includegraphics[width=1.0\textwidth]{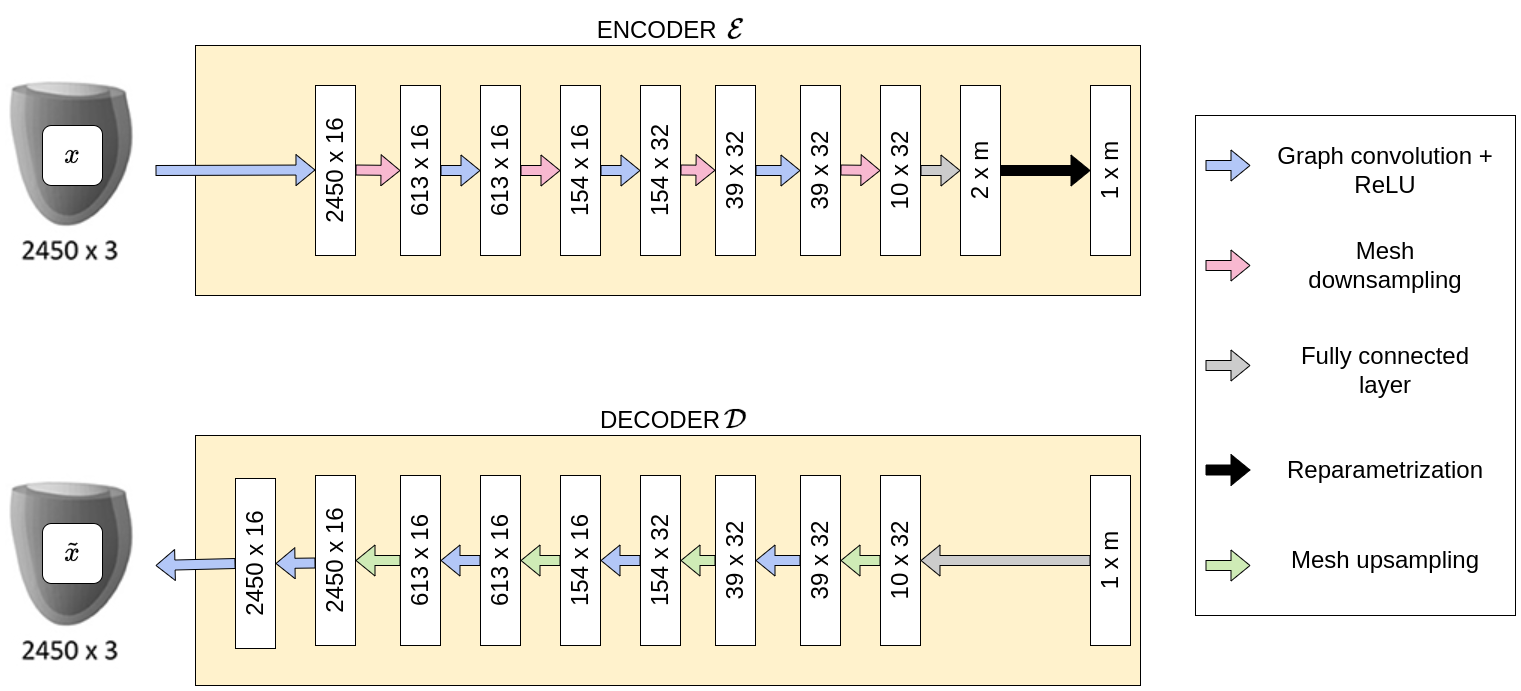}
    \caption{Illustration of the autoencoder part of the MeshLDM architecture.
    }
    \label{fig:ldm_ae}
\end{figure*}

\begin{figure*}[!t] \centering\includegraphics[width=1.0\textwidth]{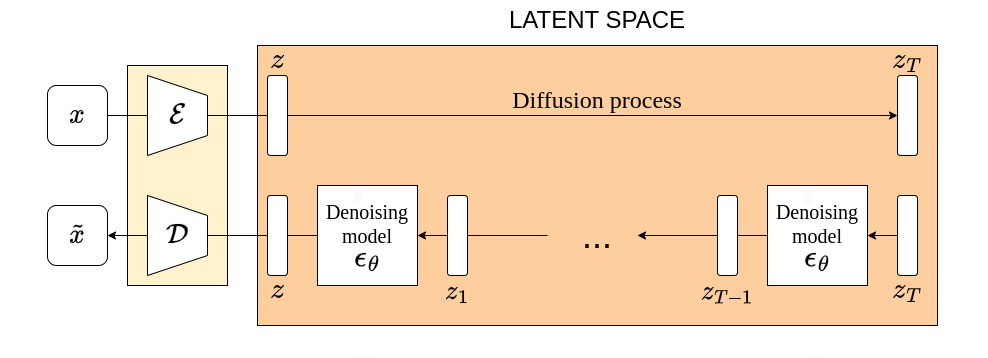}
\caption{Illustration of MeshLDM architecture. It includes an autoencoder responsible for converting data to and from latent space and a denoising model, which performs the diffusion process.}
\label{fig:ldm_dm}
\end{figure*}

The architecture of the proposed MeshLDM combines a Geometric Deep Learning \citep{bronstein2021geometric} approach with latent diffusion models \citep{rombach2021ldm}. The model includes an autoencoder, responsible for converting 3D meshes to and from latent space, and a denoising model, which performs the diffusion process. The architecture of the model is presented in Figures~\ref{fig:ldm_ae} and \ref{fig:ldm_dm}.

\subsubsection{Autoencoder}
The autoencoder is based on the Mesh VAE architecture presented in \citep{beetz2022interpretable}. It is designed to model ventricular anatomies utilizing a Geometric Deep Learning approach \citep{bronstein2021geometric}, namely graph convolutions and mesh pooling layers. An illustration of the autoencoder part of the MeshLDM architecture can be seen in Figure~\ref{fig:ldm_ae}. 
The autoencoder allows us to map the input mesh to a latent space representation, capturing its anatomical structure in a compressed form.
The dimension of the latent space, marked as \(m\) in Figure~\ref{fig:ldm_ae}, is set to 16 in our experiments, following \citet{beetz2022interpretable}. In the  training of the autoencoder, we employ Kullback-Leibler (KL) regularization. \newline

\subsubsection{Denoising model}
After encoding the entire dataset, we use it to train the denoising model. The denoising model learns to model the diffusion process in this latent space by progressively removing noise.
The original LDM \citep{rombach2021ldm} used an encoder to preserve the 2D data structure in image data and used a U-Net architecture as a denoising model.
For our MeshLDM, we use a fully connected network as the denoising model. We employ 6 fully connected layers, with each layer having both input and output dimensions of 16. Between every pair of these layers, the Swish-1 activation function \citep{ramachandran2017searching} is applied.
Additionally, at each layer, we add a diffusion timestep \(t\) encoded as a sinusoidal position embedding to the data, as described in \citep{ho2020denoising}.
As our noise scheduler, we use the DDPMScheduler introduced in \citep{ho2020denoising} and implemented in the Diffusers library \citep{2022diffusers}, with steps = 1000, \(\upbeta_0 = 0.0001\), and \(\upbeta_T = 0.02\). Specifically, steps = 1000 means the diffusion process is discretized into 1000 timesteps. \(\upbeta_0\) and \(\upbeta_T\) define the starting and ending values of the noise variance schedule. At each step, a small amount of Gaussian noise is added, with the noise level increasing from \(\upbeta_0\) to \(\upbeta_T\) linearly across the 1000 steps. These values control how gradually the data is noised during the forward diffusion process.

\subsection{Network training and implementation}
\label{s:network_training}
\subsubsection{Autoencoder}
We first train the autoencoder component of the LDM. We use the same Mesh VAE architecture as proposed in \citep{beetz2022interpretable}, where the dimension of the latent space is set to 16. We train the Mesh VAE using the same loss function for \(\upbeta\)\text{-}VAE framework and suggested parameters -- the batch size of 8, Adam optimizer \citep{optimizer} with a learning rate of 0.001, and 250 epochs. We train two separate models -- one for ED data and one for ES data.

\subsubsection{Denoising model}
After training the autoencoder component of our MeshLDM, we use it to encode the training data into the latent space. Then, we normalize the obtained vectors, which become our training dataset for the denoising model. For training the denoising model we use the batch size of 64, the Adam optimizer, and a learning rate scheduler - LambdaLR from PyTorch framework \citep{Paszke2019PyTorchAI}, initialized with a learning rate of 0.005.

We sample 1,000 random vectors (for each of the two models) from a normal distribution and input them into the trained denoising model. The resulting output vectors are scaled by the standard deviation and then shifted by the mean previously calculated during data normalization. Finally, we use the decoder to generate the corresponding synthetic 3D meshes.

\subsection{Evaluation metrics}
\label{sec:metrics}
We evaluate the MeshLDM using several clinical and 3D metrics to assess its generative capabilities.
We incorporate as clinical metrics the LV endocardial volume and LV myocardial mass.
LV endocardial volume is the volume of inner cavity of the LV, whereas LV myocardial mass is calculated as the product of the LV myocardial volume (difference between LV epicardial and LV endocardial volumes) and the density of the myocardial tissue. The average density of the myocardium is assumed to be \( \rho = 1.05~\text{g/mL} \)  \citep{beetz2022interpretable}. These metrics are commonly used in clinical practice, making them a suitable choice for evaluating the realism of the generated samples.

We additionally evaluate the generated meshes using quantitative metrics commonly employed for assessing the quality of generated 3D structures \citep{pointflow,cardiac_snowflake}. To this end, we convert the generated and reference (test dataset) meshes into point clouds by retaining only the vertices and removing the edges and faces. Let $P_g$ denote the set of point clouds derived from the generated meshes, and let $P_r$ represent the set of point clouds obtained from the reference meshes. When applying the point cloud metrics, we randomly select 285 meshes from the 1,000 generated meshes to ensure $|P_g| = |P_r|$. Here, $|P|$ denotes the number of point clouds in the dataset 
P. We use the following metrics:\newline

\textbf{Coverage (COV)} \citep{pmlr-v80-achlioptas18a}: For each point cloud in the generated set we first find the closest neighbor in the set of the reference point clouds. The Coverage score is the fraction of point clouds from reference set matched to at least one point cloud in the generated set. Formally, it is defined as:
\begin{equation}
\textnormal{COV}(P_g, P_r) = \frac{\left|\left\{\arg\min\limits_{Y \in P_r} D(X, Y) \mid X \in P_g\right\}\right|}{|P_r|},
\end{equation}
where $D(X, Y)$ is distance between two point clouds.

Essentially, for each point cloud $X$ in the generated set $P_g$, we find the closest point cloud $Y$ in the reference set $P_r$ based on a distance metric $D(X,Y)$. The ${\arg\min_{Y \in P_r} D(X, Y)}$ part identifies which reference point cloud is the closest to a given generated point cloud $X$, and hence the numerator $|\{\arg\min_{Y \in P_r} D(X, Y) \mid X \in P_g\}|$ counts the number of point clouds in the reference set that are matched to at least one point cloud in the generated set. Finally, diving by $|P_r|$ yields the fraction of point clouds in the reference set that are matched to at least one point cloud in the generated set, measuring the coverage metric.

\textbf{Minimum Matching Distance (MMD)} \citep{pmlr-v80-achlioptas18a}: 
For every point cloud $Y$ from the reference dataset $P_r$, we calculate the distance to its nearest neighbor from the generated dataset $P_g$ and then average the result. It is formally defined as follows:
\begin{equation}
\textnormal{MMD}(P_g, P_r) = \frac{1}{|P_r|} \sum\limits_{Y \in P_r} \min\limits_{X \in P_g} D(X, Y).
\end{equation}

\textbf{1-Nearest Neighbor Accuracy (1-NNA)}: It was introduced by \citet{lopez-paz2017revisiting} and later adapted by \citet{pointflow} for evaluating point cloud generative models. The 1-NNA measures the similarity between the generated and reference distributions. For each point cloud, we decide if it comes from the generated or reference dataset based on its nearest neighbor. A 1-NNA value closer to 50\% indicates higher similarity between the distributions. Let \mbox{$P_{-X} = P_r \cup P_g - \{X\}$} and $N_X$ be the nearest neighbor of \mbox{$X$ in $P_{-X}$}. Formally, 1-NNA is defined as follows:
\begin{equation}
\textnormal{1-NNA}(P_g, P_r) = 
\frac{\sum\limits_{X \in P_g} \mathbb{I}[N_X \in P_g] + \sum\limits_{Y \in P_r} \mathbb{I}[N_Y \in P_r]}{|P_g| + |P_r|},
\label{eq:nna}
\end{equation}
where $\mathbb{I}[\cdot]$ is the indicator function.

In all cases, we use the Chamfer distance (CD) as the distance metric $D(\cdot)$:
\begin{equation}
\textnormal{CD}(X, Y) = \sum\limits_{x \in X} \min\limits_{y \in Y} \|x - y\|_2^2 + \sum\limits_{y \in Y} \min\limits_{x \in X} \|x - y\|_2^2,
\label{eq:cd}
\end{equation}
where $X$ and $Y$ are two point clouds with the same number of points.
The chamfer distance from Equation~\eqref{eq:cd} is used in Equation~\eqref{eq:nna} to compute nearest neighbors between point clouds. Specifically, it is used to define the distance metric for determining $N_X$.

Additionally, we calculate the metrics for samples from the training dataset similar to \citet{pointflow}, and consider it as an upper bound since they come from the distribution that we target to simulate. We randomly select 285 samples from the training set and use them to compare the training and test set distributions.

\section{Results}
\label{s:results}
We conduct both qualitative and quantitative evaluations of the generated meshes, by comparing them to the gold standard, i.e., meshes in the test dataset.

\begin{figure*}[!t]
    \centering\includegraphics[width=1.0\textwidth]{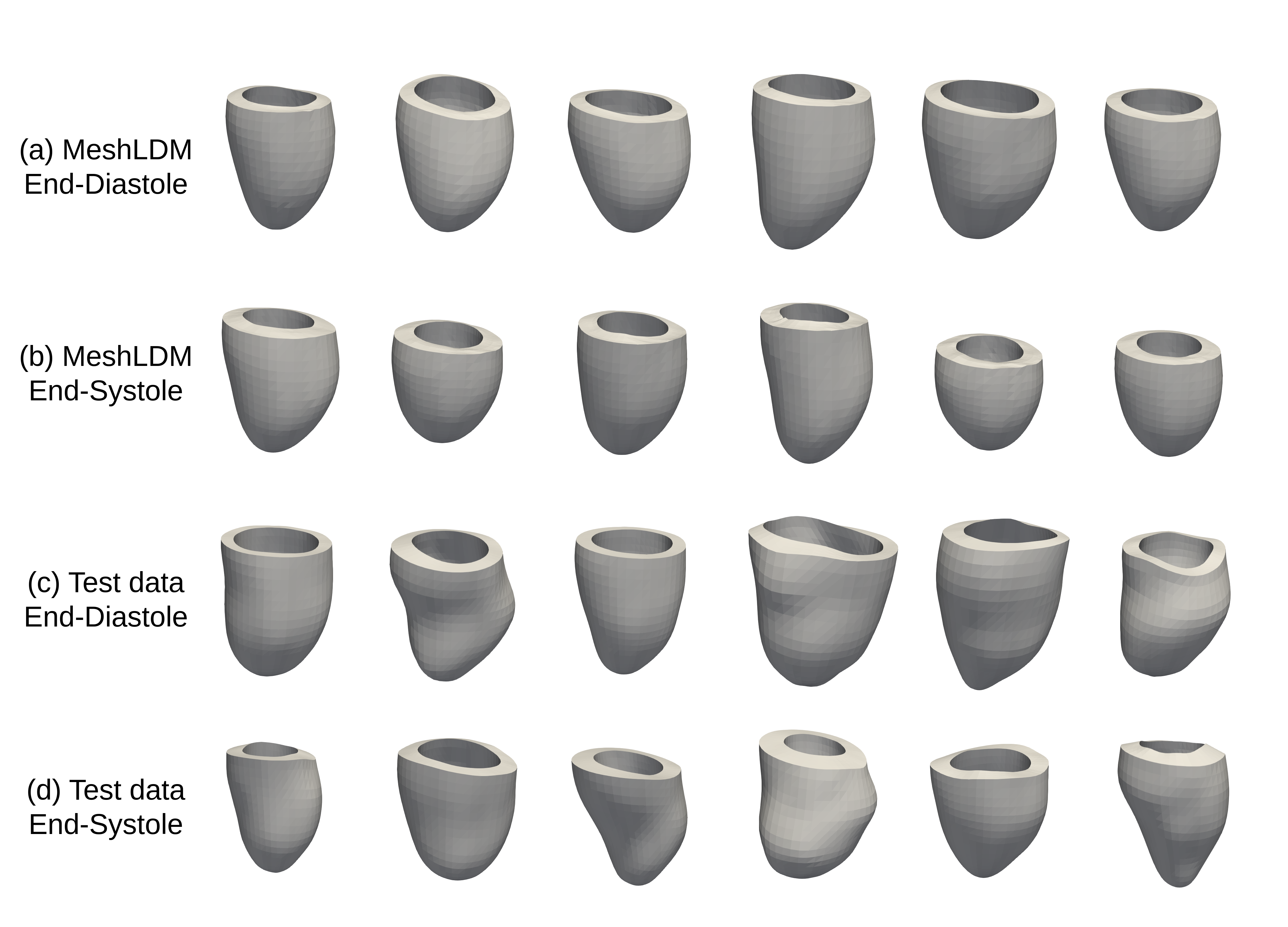}
    \caption{Sample meshes generated by MeshLDM ((a) end-diastole, (b) end-systole) and sample meshes from test dataset ((c) end-diastole, (d) end-systole).}
    \label{fig:all_meshes}
\end{figure*}

\subsection{Qualitative evaluation}
We present 6 sample cases for each phase in Figure~\ref{fig:all_meshes}. We use two separate models, one trained on the ED data and the other on the ES data. Figure~\ref{fig:all_meshes} does not present per-case correspondence between the meshes of those two cardiac phases. Similarly, the meshes in test dataset are also not in correspondence. Since we aim to generate new virtual meshes, not to reconstruct a given mesh based on a specific instance from the test dataset, pairwise comparisons between the generated and test meshes are not possible. We observe that the proposed MeshLDM successfully learns characteristics of the ED and ES meshes. The generated ED meshes have a larger overall volume and thinner myocardium than the ES meshes, which can also be observed in the meshes from the test dataset in Figure~\ref{fig:all_meshes}(c)-(d).

As shown in Figure 3(a)–(b), comparing meshes within a single row reveals notable anatomical diversity, including variations in the pointedness of the apex (e.g., meshes in row (b), columns 1–3) and differences in longitudinal curvature (e.g., meshes in row (a), columns 3–4). Additionally, the meshes generated by MeshLDM demonstrate diversity in terms of overall heart size, mid-cavity diameter, myocardial thickness, basal plane tilt, and left ventricular elongation.
However, they present less diversity than the gold standard meshes in the test dataset, as further quantified next.

\subsection{Evaluation of clinical metrics}
We calculate clinical metrics, namely LV volume and LV mass, for each mesh and present their mean and standard deviation in Table~\ref{tab:clinical_metrics}. For comparison, we provide the same metrics for the meshes in the unseen test dataset, which we consider the gold standard in this study (\mbox{$N=285$}).

We observe that MeshLDM can successfully capture the average shape of meshes for both cardiac phases, with the average difference in population mean across all scores at 2.4\%. The mean LV volume and LV mass of the meshes in ED cardiac phase are very close to the ones from the gold standard, while the means in ES cardiac phase are slightly higher. The standard deviations of the metrics in both cardiac phases are lower than the ones from the gold standard. This indicates that the MeshLDM can successfully learn the 3D shapes of the meshes, though the generated meshes represent less diversity than those from the gold standard.

\begin{table}[!t] 
  \centering
  \caption{Clinical metrics achieved by MeshLDM and MeshVAE against gold standard (test dataset).}
  \label{tab:clinical_metrics}
  \vspace{10pt} 
  \renewcommand{\arraystretch}{1.5} 
  \setlength{\tabcolsep}{4pt} 
    \begin{tabular}{|l|l|r|r|c|r|c|}
      \hline
      Phase & Metric 
      & \makecell{\\Gold standard \\[2pt] {\small[N=285]}}
      & \makecell{\\MeshLDM \\[2pt] {\small[N=1{,}000]}}  
      & \makecell{\\ Relative error \\[2pt] MeshLDM}
      & \makecell{\\MeshVAE \\[2pt]
      {\small[N=1{,}000]}}  
      & \makecell{\\ Relative error\\[2pt] MeshVAE}
      \tabularnewline
      \hline
      ED 
      & \makecell[l]{\\[-7pt] LV cavity \\[2pt] volume (ml)}
      & 156.3 \(\pm\) 43.0 & 155.8 \(\pm\) 29.2 & 0.32\% & 162.6 \(\pm\) 44.7 & 4.03\% \tabularnewline
      \cline{2-7}
      & LV mass (g) & 123.0 \(\pm\) 30.6 & 123.1 \(\pm\) 17.6 & 0.08\% & 129.6 \(\pm\) 28.7 
      & 5.37\% 
      \tabularnewline
      \hline
      ES 
      & \makecell[l]{\\[-7pt] LV cavity \\[2pt] volume (ml)}
      & 83.9 \(\pm\) 39.6 & 78.0 \(\pm\) 19.6 & 7.03\% & 84.6 \(\pm\) 28.5 
      & 0.83\%
      \tabularnewline
      \cline{2-7}
      & LV mass (g) & 130.7 \(\pm\) 33.9 & 126.9 \(\pm\) 18.0 & 2.09\% & 130.7 \(\pm\) 32.5
      & 0\%
      \tabularnewline
      \hline
      \multicolumn{5}{@{}l}{Values represent mean \(\pm\) standard deviation.}
    \end{tabular}
\end{table}

\subsection{Evaluation of cardiac structures generation}
We compute several commonly used metrics, namely 1-NNA classifier, Coverage score, and Minimum Matching Distance, on the generated meshes and training dataset. The results are summarized in Table~\ref{tab:MeshLDM_metrics}. As described in Section~\ref{sec:metrics}, the 1-NNA value should converge to 50\% if the compared distributions are the same, as observed in our results between the training and gold standard test sets. The ED samples generated by MeshLDM achieve a value of approximately 63\%, indicating a slight difference between the test and generated distributions while demonstrating a strong ability to accurately capture the variety of shapes observed in real patients' data. The ES distribution reaches around 72\% suggesting that features of ES shapes might be a bit more challenging to capture for the model. The Coverage score results indicate similar conclusions in terms of the difference between the generated and test distributions, as well as the lower performance for the ES phase. On the other hand, the minimum matching distance for the ED phase demonstrates that the generated structures are very similar to the gold standard distribution, with a difference of 13.05 mm, whereas the difference between the training set and test set is only slightly lower, at 12.86 mm.

\begin{table}[!t]
  \centering
  \caption{3D metrics for meshes generated by MeshLDM and MeshVAE against training dataset. $\uparrow$: Higher values indicate better performance; $\downarrow$: Lower values indicate better performance.}
  \label{tab:MeshLDM_metrics}
  \vspace{10pt} 
  \renewcommand{\arraystretch}{1.5} 
  \setlength{\tabcolsep}{5pt} 
    \begin{tabular}{|l|l|r|r|r|}
      \hline
      Phase & Metric & Training set & MeshLDM & MeshVAE \\
      \hline
      \multirow{3}{*}{ED} & 1-NNA (\%) & 49.48 & 63.46 & 62.06 \\
      \cline{2-5}
      & COV (\%, $\uparrow$) & 51.05 & 30.77 & 50 \\
      \cline{2-5}
      & MMD (mm, $\downarrow$) & 12.86 & 13.05 & 12.54 \\
      \hline
      \multirow{3}{*}{ES} & 1-NNA (\%) & 49.13 & 72.20 & 59.44 \\
      \cline{2-5}
      & COV(\%, $\uparrow$) & 51.75 & 23.08 & 50.7 \\
      \cline{2-5}
      & MMD (mm, $\downarrow$) & 13.48 & 14.83 & 12.96 \\
      \hline
    \end{tabular}
\end{table}

\subsection{Comparison with MeshVAE}

To enable a meaningful comparison, we include results from the Mesh Variational Autoencoder (MeshVAE) model. This model was originally proposed by \citet{beetz2022interpretable} for the modeling of ventricular anatomies, and serves as the autoencoder part of our latent diffusion model.

In terms of clinical metrics, the relative errors of the means are lower for MeshLDM for the ED phase and lower for MeshVAE for the ES phase (Table~\ref{tab:clinical_metrics}). Meanwhile, the standard deviations produced by MeshVAE are closer to those of the gold standard than those produced by MeshLDM, indicating greater capabilities to capture diversity. This is also reflected in Table~\ref{tab:MeshLDM_metrics}, with MeshVAE demonstrating a closer alignment with the training set in terms of coverage. However, the results of both methods are comparable for the rest of considered 3D metrics, especially in the case of MMD.

Altogether, MeshLDM proves to be a strong alternative to MeshVAE, though neither method consistently outperforms the other across all metrics.

\section{Discussion}
\label{s:discussion}

In this work, we have presented the first study investigating the application of LDMs for generating 3D meshes of cardiac anatomies. We have proposed a novel architecture, MeshLDM, which can generate realistic 3D meshes of LV cardiac anatomies, by combining a Geometric Deep Learning approach with diffusion models. The proposed model successfully captures the characteristics of 3D cardiac shapes and generates realistic 3D human left ventricles with only a 2.4\% difference in population mean across all clinical evaluation metrics for both cardiac phases.
The generated meshes exhibit variations in basal plane tilt, mid-cavity diameter, and shape elongation, though the overall diversity is relatively lower compared to the gold standard 3D meshes. This observation is consistent with clinical metrics and mesh generation quality, where standard deviations in both cardiac phases are lower than the ones from gold standard.

A potential solution to increase the diversity of the generated meshes would be to use a larger dataset. LDMs are known for their ability to generate diverse data, but this is only possible if the model is trained on a sufficiently large and diverse dataset. For example, Stable Diffusion was trained on LAION-400M \citep{schuhmann2021laion400mopendatasetclipfiltered}, a publicly available dataset of 400 million image-text pairs \citep{rombach2021ldm}, containing a wide variety of images and captions. In contrast, the dataset used in this study is relatively small and contains only 1,034 meshes of human left ventricles, which may have affected diversity of the generated meshes, though the size of our dataset is satisfactory in terms of post-MI 3D reconstruction.

One of the potential downstream applications of the proposed MeshLDM model is to apply it for generating synthetic cardiac structures for cardiac disease classification tasks \citep{suinesiaputra2017statistical,isensee2018automatic,bernard2018deep,zhang2019deep,avard2022non,CORRALACERO20221563,beetz2023multiobj,beetz2023post,beetz20233dshape}. In cardiology, 3D shape has been extensively studied to better understand and simulate the anatomy and function of both the healthy and diseased heart \citep{bai2015bi,gilbert2019independent,vincent2021atlas}. 3D-based biomarkers have also been shown to provide significant diagnostic value for a plethora of cardiovascular diseases, including MI \citep{suinesiaputra2017statistical,CORRALACERO20221563}. While traditional machine learning approaches to 3D cardiac shape analysis have primarily employed principal component analysis and related methods, more recent work has increasingly focused on deep learning techniques \citep{gilbert2020artificial}. While early deep learning approaches were based on grid-based representations of 3D shape \citep{qin2018joint,xu2019ventricle,biffi2020explainable}, newer studies have also explored geometric deep learning applied to both point cloud \citep{beetz2021generating,beetz20243dcontraction,cardiac_snowflake,seale2024multiphase} and mesh representations of the heart \citep{kong2021deep,chen2021shape,beetz2023mesh,meng2023deepmesh,kalaie2024geometric}. These existing approaches typically depend on large and high-quality cardiac data which is time-consuming, costly, and difficult to acquire. Our proposed MeshLDM can mitigate these challenges by synthesizing high-resolution, diverse, and realistic 3D cardiac structures.
The generated synthetic data have valuable applications, including use in electrophysiological computer simulations for \textit{in silico} trials \mbox{\citep{li2024towards,Coleman2024,Dasi2024,CAMPS2025103361}} and data augmentation for training cardiac disease classifiers \mbox{\citep{beetz2023post,CORRALACERO20221563,CorralAcero2024}}. However, demonstrating these downstream uses is well beyond the scope of this manuscript, as it would necessitate the development, training, and validation of new application-specific models and evaluation frameworks, each representing a significant research effort beyond the current study.

Our work is an important step towards exploring the possibilities of diffusion-based models in the field of medical imaging, especially 3D anatomy generation. 
As a future extension, a conditioning mechanism can be incorporated in the MeshLDM that would allow us to generate data for specific subpopulations using the same model, instead of training separate models for different subgroups such as cardiac phases.
Such a conditioning mechanism would enable, for instance, to train a single model on both ED and ES data. While incorporating a conditioning mechanism is a natural and promising extension, it was outside the scope of the current study. We focused on demonstrating the possibility of using latent diffusion models to generate realistic 3D cardiac anatomies for individual phases.

LV volume and mass are important clinical indicators of cardiac health, and LV volume is widely used in current clinical practice. Although they do not provide a detailed characterization of cardiac shapes, they help indicate whether the generated meshes fall within a plausible clinical range. As part of our future work, once the aforementioned conditioning mechanism enabling the correspondence between the ES and ED phases is implemented, further metrics such as LV ejection fraction could be computed to provide more comprehensive comparisons with established clinical metrics.

Further future work on the automatization process of cardiac mesh generation can also incorporate intermediate temporal phases (along with ED and ES phases) for the modeling of cardiac mechanics, as well as diverse demographic characteristics such as sex, age, body mass index, etc., towards subpopulation-specific cardiac anatomy modeling \citep{beetz2021generating}.

\section{Conclusion}
In this study, we have introduced MeshLDM, a novel latent diffusion model architecture specifically designed for generating 3D meshes of human cardiac anatomies. Our extensive evaluations demonstrate that MeshLDM effectively captures the intricate characteristics of human cardiac anatomy at both end-diastolic and end-systolic phases across a wide range of clinical and 3D mesh reconstruction metrics, although with slightly less diversity compared to the clinical data. Nonetheless, our work represents a pioneering application of LDMs in 3D cardiac anatomy generation, offering a promising tool for addressing data scarcity in medical imaging and enhancing the realism and diversity of synthetic cardiac data. The successful implementation of MeshLDM underscores the potential impact of this and other generative models on advancing cardiac disease modeling, diagnosis, and treatment planning through high-fidelity virtual anatomy generation.

\section*{Conflict of Interest Statement}
The authors declare that the research was conducted in the absence of any commercial or financial relationships that could be construed as a potential conflict of interest.

\section*{Author contributions}
JM: Conceptualization, Formal Analysis, Investigation, Methodology, Software, Visualisation, Writing - original draft, Writing – review \& editing.
MB: Investigation, Data Curation, Writing – review \& editing.
LMK: Investigation, Writing – review \& editing.
VG: Investigation, Methodology, Project Administration, Data Curation, Writing – review \& editing, Supervision.
AB: Conceptualization, Investigation, Methodology, Project Administration, Writing – review \& editing, Supervision, Funding acquisition.
ABO: Conceptualization, Investigation, Methodology, Project Administration, Data Curation, Writing – review \& editing, Supervision, Funding acquisition.

\section*{Funding}
A.~Banerjee holds a Royal Society University Research Fellowship (Grant No.~URF{\textbackslash}R1{\textbackslash}221314). A.~Bueno-Orovio acknowledges support from the Innovate UK grant 10110728.

\section*{Acknowledgments}
This work made use of the University of Oxford's Advanced Research Computing (ARC) facility (\url{https://doi.org/10.5281/zenodo.22558}).

\section*{Code and data availability}
The code is available at \url{https://github.com/mozyrska/Mesh-LDM}. The datasets used in this article is not publicly available due to ethical and legal restrictions related to patient confidentiality. Generated virtual data can be made available upon reasonable request.


\bibliographystyle{elsarticle-harv} 
\bibliography{cas-refs}



\end{document}